\documentclass[prb,twocolumn,showpacs,floatfix,superscriptaddress]{revtex4-1}

\usepackage{amssymb,amsmath,amstext}                
\usepackage{graphicx}                                               
\usepackage{epstopdf}                                               
\usepackage{color}       
\usepackage{bm}
\usepackage{appendix}                                              
\usepackage[utf8]{inputenc}
\usepackage{bbold}
\usepackage{bbm}
\usepackage{latexsym}
\usepackage{mathrsfs}
\usepackage[colorlinks=true,citecolor=blue,linkcolor=magenta]{hyperref}

\newcommand{\pc}[1]{{\color{blue}{[PC: {#1}}]}}

\def\be{\begin{equation}}
\def\ee{\end{equation}}

\def\ba{\begin{eqnarray}}
\def\ea{\end{eqnarray}}

\begin{document}
\title{Tensor network simulation of the Kitaev-Heisenberg model at finite temperature}

\author{Piotr Czarnik}
\affiliation{Institute of Nuclear Physics, Polish Academy of Sciences, Radzikowskiego 152, PL-31342 Krak\'ow, Poland}

\author{Anna Francuz}
\affiliation{ Marian Smoluchowski Institute of Physics, Jagiellonian University,
              ul. Prof. S. {\L}ojasiewicza 11, PL-30-348 Krak\'ow, Poland }   
              
\author{Jacek Dziarmaga}
\affiliation{ Marian Smoluchowski Institute of Physics, Jagiellonian University,
              ul. Prof. S. {\L}ojasiewicza 11, PL-30-348 Krak\'ow, Poland }
              
\begin{abstract}  
We investigate the Kitaev-Heisenberg (KH) model at finite temperature using the exact environment full update (eeFU), introduced in Phys. Rev. B 99, 035115 (2019), which represents purification of a thermal density matrix on an infinite hexagonal lattice by an infinite projected entangled pair state (iPEPS). 
We show that, thanks to a dynamical mapping from a hexagonal to a rhombic lattice,
the eeFU on the hexagonal lattice is as efficient as the simple full update (FU) algorithm. 
Critical temperatures for coupling constants in the stripy and the antiferromagnetic phase are estimated. 
They are an order of magnitude less than the couplings in the Hamiltonian. 
By a duality transformation,
these results can be mapped to, respectively, the ferromagnetic and zigzag phases. 
For the special case of the pure Kitaev model,
which is tractable by quantum Monte-Carlo but the most challenging for tensor networks, 
the algorithm is benchmarked against the Monte-Carlo results. 
It recovers accurately the crossover to spin ordering and qualitatively the one to flux ordering.
\end{abstract}
\maketitle

\section{Introduction}
\label{sec:introduction}

Weakly entangled quantum states can be efficiently represented by tensor networks \cite{Verstraete_review_08,Orus_review_14}: either a 1D matrix product state (MPS) \cite{Fannes_MPS_92}, its 2D generalization to a projected entangled pair state (PEPS) \cite{Verstraete_PEPS_04}, or a multi-scale entanglement renormalization ansatz (MERA) \cite{Vidal_MERA_07,Vidal_MERA_08,Evenbly_branchMERA_14,Evenbly_branchMERAarea_14}.
The MPS is a compact representation of ground states of 1D gapped local Hamiltonians \cite{Verstraete_review_08,Hastings_GSarealaw_07,Schuch_MPSapprox_08} and purifications of their thermal states \cite{Barthel_1DTMPSapprox_17}. It is also the ansatz underlying the powerful density matrix renormalization group (DMRG) \cite{White_DMRG_92, White_DMRG_93,Schollwock_review_05,Schollwock_review_11}. Analogously, the 2D PEPS is expected to represent  ground states of 2D gapped local Hamiltonians \cite{Verstraete_review_08,Orus_review_14} and their thermal states \cite{Wolf_Tarealaw_08,Molnar_TPEPSapprox_15}, though representability of area-law states in general was shown to have its limitations \cite{Eisert_TNapprox_16}. Tensor networks evade the sign problem plaguing the quantum Monte Carlo, hence they can deal with fermionic systems \cite{Corboz_fMERA_10,Eisert_fMERA_09,Corboz_fMERA_09,Barthel_fTN_09,Gu_fTN_10}, as was shown for both finite \cite{Cirac_fPEPS_10} and infinite PEPS \cite{Corboz_fiPEPS_10,Corboz_stripes_11}.

The PEPS was proposed originally for ground states of finite systems \cite{Verstraete_PEPS_04,Murg_finitePEPS_07} generalizing earlier attempts to construct trial wave-functions for specific models \cite{Nishino_2DvarTN_04}. Efficient numerical methods for infinite PEPS (iPEPS) \cite{Cirac_iPEPS_08,Xiang_SU_08,Gu_TERG_08,Orus_CTM_09} promoted it to a versatile tool for strongly correlated systems in 2D. Examples of their potential include a solution of the long standing magnetization plateaus problem in the highly frustrated compound $\textrm{SrCu}_2(\textrm{BO}_3)_2$ \cite{Matsuda_SS_13,Corboz_SS_14}, demonstration of the striped nature of the ground state of the doped 2D Hubbard model \cite{Simons_Hubb_17}, and new evidence supporting gapless spin liquid (SL) in the kagome Heisenberg antiferromagnet \cite{Xinag_kagome_17}. Recent developments in iPEPS optimization \cite{fu,Corboz_varopt_16,Vanderstraeten_varopt_16}, contraction \cite{Fishman_FPCTM_17,Xie_PEPScontr_17}, energy extrapolations \cite{Corboz_Eextrap_16}, and universality class estimation \cite{Corboz_FCLS_18,Rader_FCLS_18,Rams_xiD_18} open possibility of applying it to even more difficult problems, including simulation of thermal states \cite{Czarnik_evproj_12,Czarnik_fevproj_14,Czarnik_SCevproj_15, Czarnik_compass_16,Czarnik_VTNR_15,Czarnik_fVTNR_16,Czarnik_eg_17,Dai_fidelity_17,CzarnikDziarmagaCorboz,CzarnikCorbozXiD,Orus_SUfiniteT_18}, 
mixed states of open systems \cite{Kshetrimayum_diss_17,CzarnikDziarmagaCorboz}, 
exited states \cite{Vanderstraeten_tangentPEPS_15},
or real time evolution of 2D quantum states \cite{CzarnikDziarmagaCorboz,HubigCirac} .

In parallel with iPEPS, progress was made in simulating systems on cylinders of finite circumference with DMRG.
This method of high numerical stability is routinely used to investigate 2D ground states \cite{Simons_Hubb_17,CincioVidal} and recently was applied also to thermal states on a cylinder \cite{Stoudenmire_2DMETTS_17,Weichselbaum_Tdec_18,WeichselbaumTriangular,WeichselbaumBenchmark} but the exponential growth of the bond dimension limits the circumference to a few lattice sites. Among alternative approaches are methods of direct contraction and renormalization of a 3D tensor network representing a 2D thermal density matrix \cite{Li_LTRG_11,Xie_HOSRG_12,Ran_ODTNS_12,Ran_NCD_13,Ran_THAFstar_18,Su_THAFoctakagome_17,Su_THAFkagome_17,Ran_Tembedding_18}.

\begin{figure}[t]
\includegraphics[width=0.7\columnwidth,clip=true]{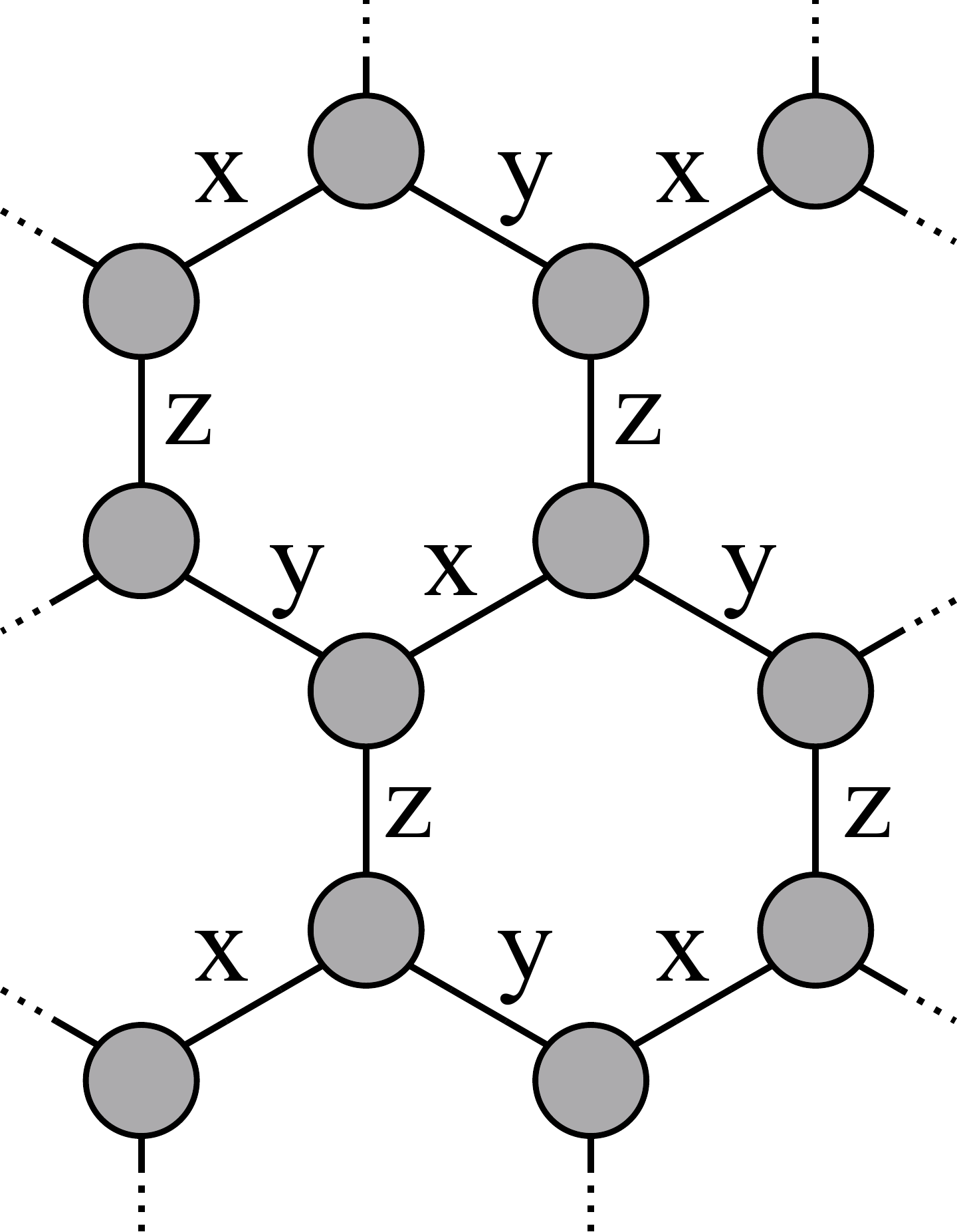}
\vspace{-0cm}
\caption{
Infinite hexagonal pseudospin-$1/2$ lattice with bonds $\gamma=x,y,z$. 
In the Kitaev model there are Ising-like couplings between $\gamma$-components of 
nearest-neighbor pseudospins connected by a $\gamma$-bond. 
}     
\label{fig:hex}
\end{figure}

The Kitaev model \cite{KitaevModel} is an exactly solvable pseudospin-$1/2$ system on a hexagonal lattice with Ising-like couplings of $\gamma=x,y,z$ components of nearest-neighbor pseudospins with strength $K_\gamma$ along $\gamma$-bonds, see Fig. \ref{fig:hex}. In any of its three A-phases, when one of the three couplings dominates, the model reduces to the effective toric code Hamiltonian \cite{ToricCode}. On the other hand, in the B-phase, where the three couplings are of similar strength, the ground state of the highly frustrated model is a critical quantum spin liquid (SL). In the SL a magnetic field opens a finite energy gap protecting a chiral topological order. Its non-abelian anyonic excitations can be employed to perform universal topological quantum computation \cite{TopoComp}. This motivates intensive search for a robust physical implementation of the model. 

In spin systems their $SU(2)$ symmetry constrains the interaction to be of the Heisenberg type. In order to break the symmetry -- and introduce the bond-anisotropy at the same time -- the spins can be mixed with orbital degrees of freedom, as originally argued for iridium oxides \cite{Chaloupka2010}. The resulting Kitaev-Heisenberg (KH) model was considered as a minimal model to study stability of the Kitaev spin liquid phase in materials like $\textrm{Na}_2\textrm{IrO}_3$, $\alpha\textrm{-Li}_2\textrm{IrO}_3$, $\textrm{Li}_2\textrm{RhO}_3$, and $\alpha\textrm{-RuCl}_3$, though recent results suggest that more general extensions of the KH model are required\cite{ValentiReview,Gotfryd2019}. In this paper we follow Refs. \onlinecite{Chaloupka2010,KHfiniteT,Chaloupka2013,KHMC_PhysRevL,Nasu2D} 
and investigate the basic KH model at finite temperature as a first step towards its more realistic extensions.

The paper is organized as follows. In section \ref{model} we review the KH model and its phase diagram at zero temperature. In section \ref{TN} we briefly outline the tensor network method to simulate thermal states of a quantum Hamiltonian. Here we introduce the dynamical mapping from a hexagonal to rhombic lattice that renders the exact environment full update (eeFU) as efficient as the simplified full update (FU) algorithm.
In section \ref{estimation} a scaling theory is discussed that is necessary to extrapolate results obtained with a finite symmetry-breaking bias to zero bias field.
In section \ref{results} we present results for the striped and antiferromagnetic phases of the KH model. In section \ref{secKitaev} we benchmark our method against quantum Monte-Carlo results\cite{Nasu2D} in the pure Kitaev model with a quantum spin-liquid ground state. We conclude in section \ref{conclusion}.

\begin{figure}[t]
\includegraphics[width=\columnwidth,clip=true]{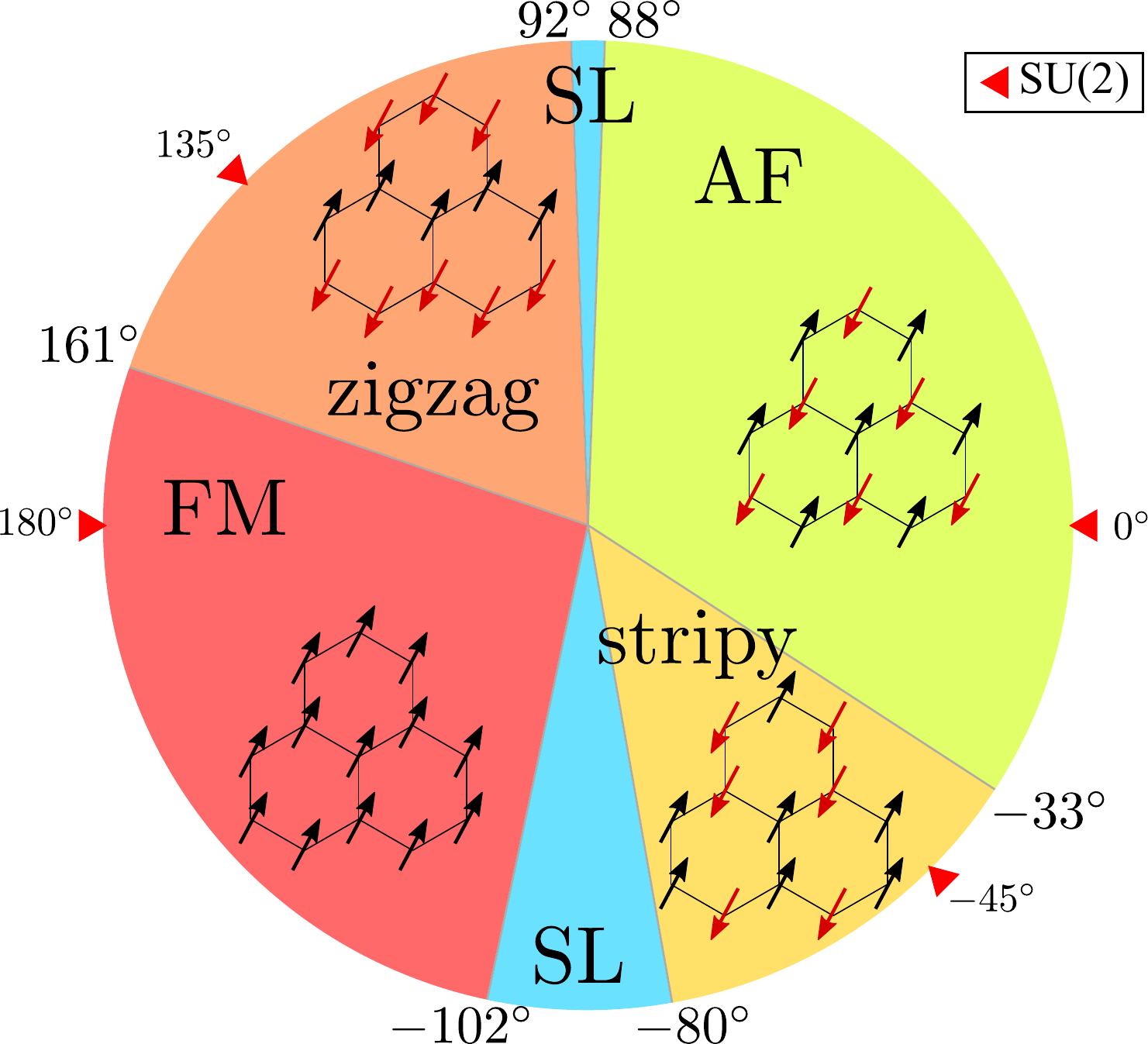}
\vspace{-0cm}
\caption{
The ground state phase diagram of the Kitaev-Heisenberg model 
\cite{Chaloupka2010,Chaloupka2013,Gotfryd2017,Corboz_KH_14}
parameterized by the angle $\phi$. The two Kitaev points, $\phi=\pm90^\circ$, are surrounded by areas of a gapless quantum spin liquid (SL). The antiferromagnetic (AF) and stripy phases on the right are connected by the duality transformation (\ref{duality}) to, respectively, the zigzag and ferromagnetic (FM) phases on the left. By the Mermin-Wagner theorem, at the four $SU(2)$-symmetric points, $\phi=-45^\circ,0^\circ,135^\circ,180^\circ$, the ordering is possible only at $T=0$. The ordered pseudospins define the order parameter pointing along either $\pm x$, $\pm y$, or $\pm z$.
}     
\label{fig:gs}
\end{figure}

\section{Model} 
\label{model}

The model \cite{Chaloupka2010,Chaloupka2013,Gotfryd2017} is a sum of nearest-neighbor terms on a hexagonal lattice,
\begin{equation}
H_{\rm KH} = \sum_{<i,j>} H_{i,j}^{(\gamma)},
\label{HKH}
\end{equation}
where 
\begin{equation}
H_{i,j}^{(\gamma)} =  J \mathbf{S}_i\cdot\mathbf{S}_j + K S_i^{\gamma} S_j^{\gamma} 
\end{equation}
depends on bond direction $\gamma= x,y,z$, see Fig. \ref{fig:hex}.
Here $\mathbf{S} = (S^x , S^y, S^z) = \frac{1}{2} (\sigma_x , \sigma_y, \sigma_z)$ are spin-$1/2$ operators defined by Pauli matrices.
The coupling constants are parameterized by an angle $\phi$:
\begin{equation}
J = A \textrm{cos}\,\phi, \quad K = 2A\,\textrm{sin}\,\phi. 
\end{equation}
Here $A>0$ is a constant. 

The zero-temperature phase diagram is shown in Fig.~\ref{fig:gs}. It was obtained by a variety of methods \cite{Chaloupka2010,Chaloupka2013,Gotfryd2017} and corroborated by iPEPS \cite{Corboz_KH_14}. Pairs of angles, $\phi$ and $\tilde\phi$ satisfying 
\be 
\tan\tilde \phi = -\tan \phi -1, 
\label{duality}
\ee  
on the right and left of the diagram, respectively, are related by a duality transformation \cite{Chaloupka2010}. There are two self-dual points, $\phi=\tilde\phi=\pm90^{\circ}$, where the model reduces to the  Kitaev model. Each of these two Kitaev points is surrounded by a gapless quantum spin liquid (SL). The same duality maps the antiferromagnetic (AF) and stripy phases on the right to, respectively, the zigzag and ferromagnetic (FM) phases on the left.

For $\phi= 0^{\circ}$ and $180^{\circ}$ the model reduces to the antiferromagnetic and ferromagnetic Heisenberg model, respectively. By the Mermin-Wagner theorem, its $\textrm{SU}(2)$-symmetry prevents spontaneous symmetry breaking at any $T>0$. The duality transformation maps these two points to $\phi=135^\circ$ and $-45^\circ$, respectively. Their hidden $SU(2)$ symmetry also prevents the ordering at finite $T$.

The frustrated model is not tractable by quantum Monte Carlo, except the pure Kitaev model \cite{Nasu2D}. Its mean-field theory is $SU(2)$-symmetric \cite{Chaloupka2013,Gotfryd2017} suggesting no finite-$T$ ordering at any $\phi$ but a spin-wave expansion and plaquette mean-field suggest a disorder-induced-order at low temperatures stabilized by both quantum and thermal fluctuations \cite{Chaloupka2013,KHfiniteT}. The latter effect is confirmed by classical Monte Carlo simulations \cite{KHMC_PhysRevL,KHMC_PhysRevB}. The model is also tractable by a high-$T$ series expansion \cite{KHseries}.  

In this work we treat the finite-$T$ KH model with a quantum tensor network for the first time. Previously we used quantum tensor networks to simulate the closely related compass and $e_g$ models \cite{Czarnik_compass_16,Czarnik_eg_17} at finite $T$ achieving good accuracy. In order to simulate the model in neighbourhood of non-analytic critical points efficiently, we add a tiny symmetry breaking bias,
\be 
H=H_{\rm KH}-\sum_i h_i S^z_i,
\label{bias}
\ee 
with a magnitude $h=|h_i|$. $h_i=h$ in the FM phase and is staggered in the AF phase. To obtain critical properties of the Kitaev-Heisenberg model we extrapolate to $h=0$ as described in Sec. \ref{estimation}.

\begin{figure}[t]
\includegraphics[width=\columnwidth,clip=true]{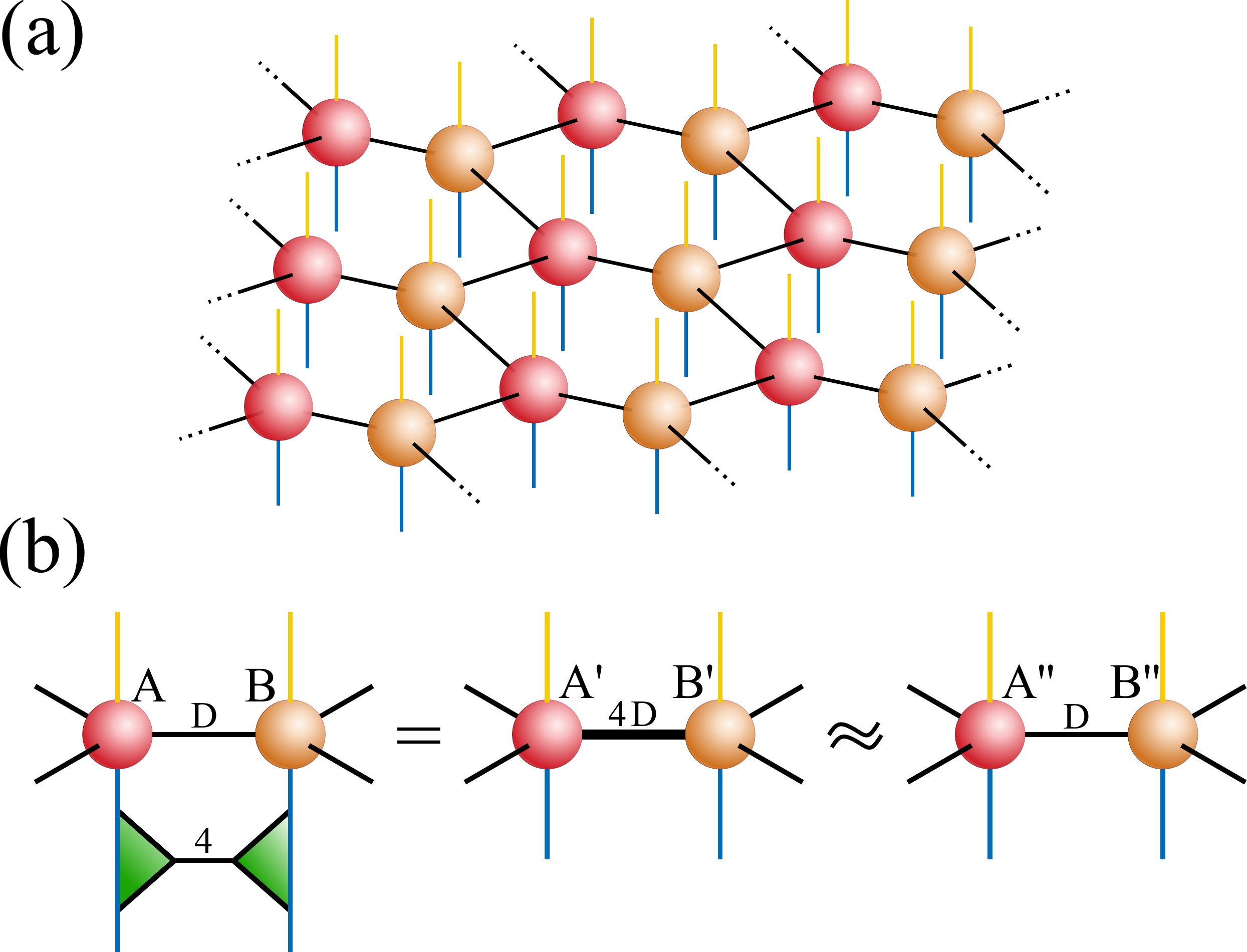}
\vspace{-0cm}
\caption{
In (a),
the iPEPS tensor network representing a purification of the thermal Gibbs state $e^{-\beta H}$. Here $\beta = 1/T$ is an inverse temperature. The pseudospin and ancilla indices are pointing down and up, respectively. The purification has two sublattices $A$ and $B$ denoted by the red and orange tensors, respectively. Nearest-neighbor tensors are contracted through bond indices with a bond dimension $D$.
In (b),
the gate $\exp(-d\beta H^{(\gamma)}_{ij}/2)$ is applied to pseudospin indices on a nearest neighbor bond $\gamma$. Here the gate is singular-value-decomposed into a contraction of two gate tensors (the green ones) connected by an index of dimension $4$ equal to the rank of the SVD. A contraction of the tensor $A$($B$) with the left(right) gate tensor yields an exact tensor $A'$($B'$). $A'$ is contracted with $B'$ by a bond index of dimension $4D$. Then the exact contraction $A'-B'$ is approximated by a contraction of new tensors $A''$, $B''$ with the original bond dimension $D$. The new tensors are optimized to minimize the error of the whole purification.
}     
\label{fig:ipeps}
\end{figure}

\section{Tensor network} 
\label{TN}
\begin{figure}[t]
\includegraphics[width=\columnwidth,clip=true]{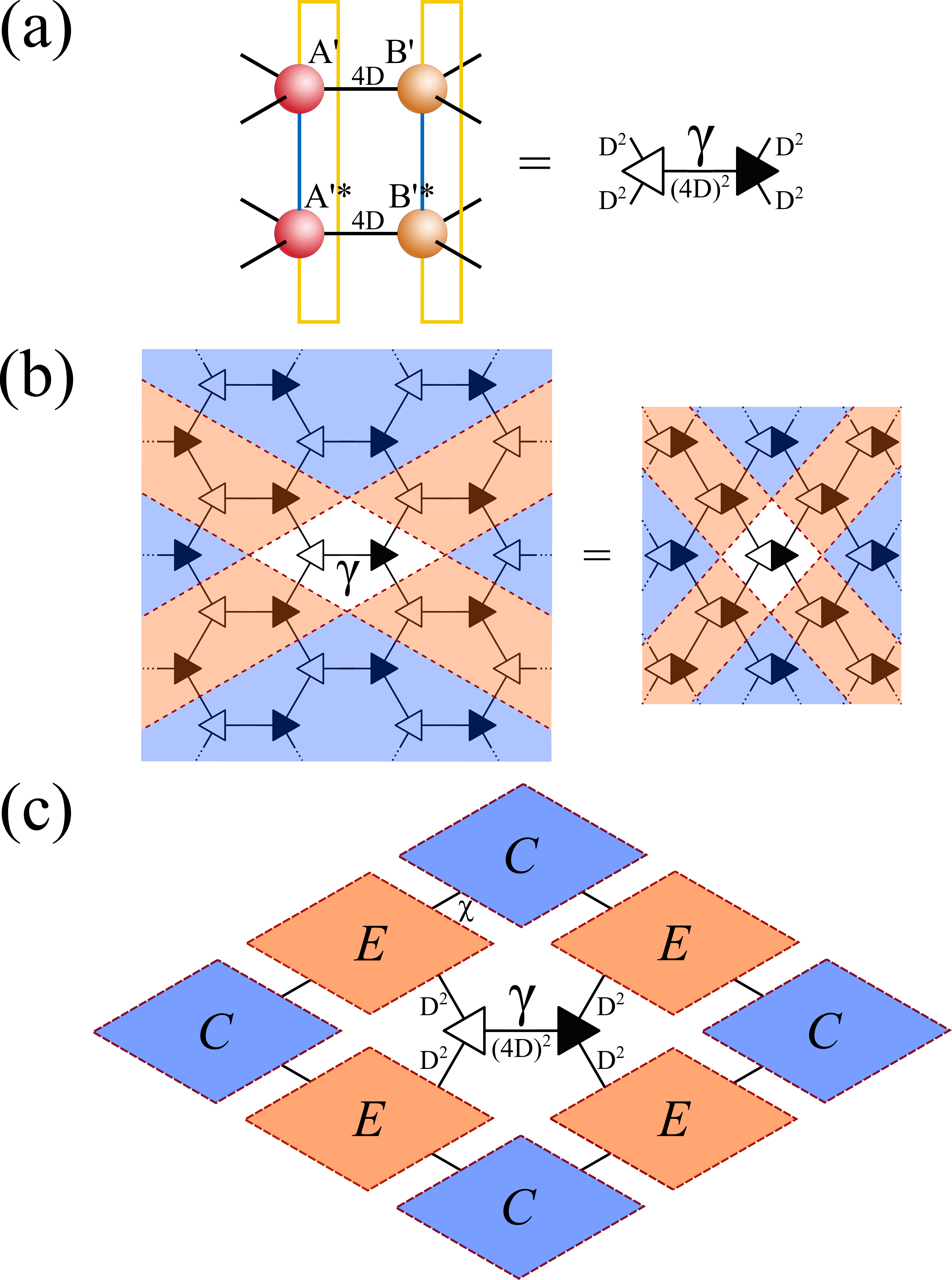}\\
\vspace*{0.5cm}
\caption{ 
Corner transfer matrix renormalization group (CTMRG).
In (a),
in a tensor network representing the norm $\langle\psi(\beta)|\psi(\beta)\rangle$ of the purification,
every iPEPS tensor $A'$ ($B'$) in the ket layer is contracted with its conjugate $A'^*$ ($B'^*$) in the bra layer to make a double tensor denoted by a white (black) triangle. The double tensors have their bond dimensions equal to either $(4D)^2$ on their $\gamma$-bonds or $D^2$ otherwise. 
In (b),
with the double tensors the norm becomes the network on the left hand side. In order to avoid handling the large dimension of the $\gamma$-bonds, on the right hand side pairs of white and black double tensors are contracted together into quadruple tensors with all bond dimensions equal $D^2$. These tensors form a rhombic lattice which is equivalent to a square lattice. 
In (c),
CTMRG is a procedure to replace the semi-infinite sectors in panel (b) by corresponding finite corner (C) and edge (E) tensors connected by indices of dimension $\chi$.  
}
\label{fig:ctm}
\end{figure}

In this work we apply the exact environment full update (eeFU) introduced and benchmarked in Ref. \onlinecite{CzarnikDziarmagaCorboz}. Here we just outline the algorithm emphasizing its adjustments to the KH model referring for more details to Ref. \onlinecite{CzarnikDziarmagaCorboz}. The most important development is the dynamical mapping from a hexagonal to rhombic lattice that makes the eeFU as efficient as the FU algorithm.

Thanks to the duality transformation (\ref{duality}), it is enough to consider the AF and FM phases only. They require only two sublattices: $A$ and $B$. We enlarge the Hilbert space by accompanying every pseudospin-$1/2$ with a pseudospin-$1/2$ ancilla. The iPEPS tensor network in Fig. \ref{fig:ipeps}(a) represents a thermal state's purification $|\psi(\beta)\rangle$ in the enlarged space. Here $\beta = 1/T$ is an inverse temperature. Its partial trace over the ancillas ($a$) yields the Gibbs state as a thermal density matrix:
\be 
{\rm Tr}_a |\psi(\beta)\rangle\langle\psi(\beta)| \propto e^{-\beta H}.
\ee 
The purification is evolved in the imaginary time $\beta$ with the eeFU algorithm:
$|\psi(\beta)\rangle=e^{-\beta H/2}|\psi(0)\rangle$.

The time evolution is represented by a product of $N$ small time steps $e^{-\beta H/2}=\left(e^{-d\beta H/2}\right)^N$, where $N=\beta/d\beta$. Each time step is subject to a second order Suzuki-Trotter decomposition \cite{Suzuki_66,Suzuki_76,Trotter_59}:
\ba
&& e^{-d\beta H/2} \approx \nonumber \\
&& G_x(d\beta/2) G_y(d\beta/2) G_z(d\beta) G_y(d\beta/2) G_x(d\beta/2),
\ea
where $G_\gamma(d\beta)=\prod_{\langle i,j \rangle\| \gamma} e^{-\frac12 d\beta H^{(\gamma)}_{ij}}$ is a product of nearest neighbor gates over all $\gamma$-bonds. Here $H^{(z)}_{ij}$ includes also the bias in Eq. (\ref{bias}). 

The action of $G_\gamma$ on one of the $\gamma$-bonds is shown in Fig. \ref{fig:ipeps}(b). A contraction of the ``old'' tensors $A,B$ with the gate $e^{-\frac12 d\beta H^{(\gamma)}_{ij}}$ becomes a contraction of new exact tensors $A'$, $B'$ with an enlarged $\gamma$-bond dimension $4D$. The exact contraction $A'-B'$ is approximated by a contraction of new tensors $A''-B''$ with the original bond dimension $D$. The new tensors are optimized to minimize the error introduced by this approximation to the whole purification. 

In order to minimize the error of the infinite purification, we need a tensor environment for the considered bond $\gamma$. To this end we treat the exact two-site contraction $A'-B'$ as if it were a single iPEPS tensor on the two sites, see Fig. \ref{fig:ctm}. Effectively, every two nearest neighbor sites connected by a $\gamma$-bond are fused into a single site. The hexagonal lattice is replaced by a rhombic one, which can be treated as a square lattice. This way we can employ full potential of the robust square-lattice corner transfer matrix renormalization group \cite{Baxter_CTM_78,Nishino_CTMRG_96,Orus_CTM_09,Corboz_CTM_14} (CTMRG) to obtain the environment for the $\gamma$-bond, see Fig. \ref{fig:ctm}(c). However, the main advantage is that every enlarged $4D$-dimensional $\gamma$-bond index is hidden inside the square-lattice composite iPEPS tensor and, hence, it does not slow down the CTMRG which is the main bottleneck of the whole algorithm. 

\section{Estimation of critical temperature}
\label{estimation}

The evolution near a critical point is challenging \cite{Czarnik_evproj_12,CzarnikDziarmagaCorboz}. In particular, finite $\chi$ limits the correlation length which can be obtained by CTMRG \cite{Nishino_scalCTM_96}, hence a large $\chi$ is necessary to render the environment of the $\gamma$-bond accurate enough to obtain correct new tensors $A''$ and $B''$. Therefore, in Refs. \onlinecite{Czarnik_evproj_12,CzarnikDziarmagaCorboz} a small symmetry-breaking bias $h$ was introduced to turn the transition into a smooth crossover making the correlation length finite and allowing for results well converged in $\chi$. However, in order to estimate $T_c$ an extrapolation back to $h=0$ was necessary. To this end, a systematic scaling theory was used \cite{CzarnikDziarmagaCorboz} yielding very accurate results for the quantum Ising model. Here we follow the same approach. 

According to the scaling theory the order parameter $m$, its derivative with respect to $T$, and the correlation length $\xi$ satisfy the scaling laws:
\ba
m(t,h)  &=&  h^{1/\delta} f(t h^{-1/\tilde\beta\delta}),
\label{mscal} \\
m'(t,h) &=&  h^{(\tilde\beta-1)/\tilde\beta\delta} f'(t h^{-1/\tilde\beta\delta}),
\label{mderivscal} \\
\xi(t,h) &=&  h^{-\nu/\tilde\beta\delta} g(t h^{-1/\tilde\beta\delta}),
\label{xiscal} \\
C_V(t,h) &=& h^{-\alpha/\tilde\beta\delta} h(t h^{-1/\tilde\beta\delta}),
\ea
respectively. Here $t=T-T_c$, the prime is a derivative with respect to $t$, $f(x)$, $g(x)$ and $h(x)$ are non-universal scaling functions, while $\tilde\beta,\delta,\nu,\alpha$ are universal critical exponents. In order to estimate $T_c$ we use an observation that, for a fixed $h$, the slope $m'(t,h)$ has a peak at $t^*=T^*-T_c>0$. In the regime of small $h$ its position, determined by the maximum $x^*$ of the scaling function $f'(x)$, should scale as 
\be  
T^*(h) = T_c + x^* h^{1/\tilde\beta\delta}. 
\label{tpeak}
\ee
Fitting numerical data for the pseudo-critical temperature, $T^*(h)$, with the function on the right hand side we estimate three parameters: $x^*$, $1/\tilde\beta\delta$ and, most importantly, $T_c$.
Similarly we observe that also $C_V(t,h)$ has a maximum at $\tilde{t}^*=\tilde{T}^*-T_c>0$, which position scales the same as $T^*(h)$ in (\ref{tpeak}) making it also suitable to estimate $1/\tilde\beta\delta$ and $T_c$.

Furthermore, we use the behavior of $\xi(t^*,h)$ and $m'(t^*,h)$ to test self-consistency of the scaling theory. We observe that $\xi(t^*,h)$ is close to the maximal correlation length for a given bias $h$ and $m'(t^*,h)$ is the maximal magnetization's slope by definition. Equations (\ref{mderivscal},\ref{xiscal},\ref{tpeak}) imply two power laws that do not depend on the unknown $T_c$: 
\be
m'(t^*,h)  \propto  h^{(\tilde\beta-1)/\tilde\beta\delta}, 
\label{m'peak}
\ee
\be
\xi(t^*,h) \propto  h^{-\nu/\tilde\beta\delta}.
\label{xipeak}
\ee
Therefore, they provide a reliable test whether $h$ is small enough to achieve the critical scaling regime.

\begin{figure}[h]
\includegraphics[width=\columnwidth,clip=true]{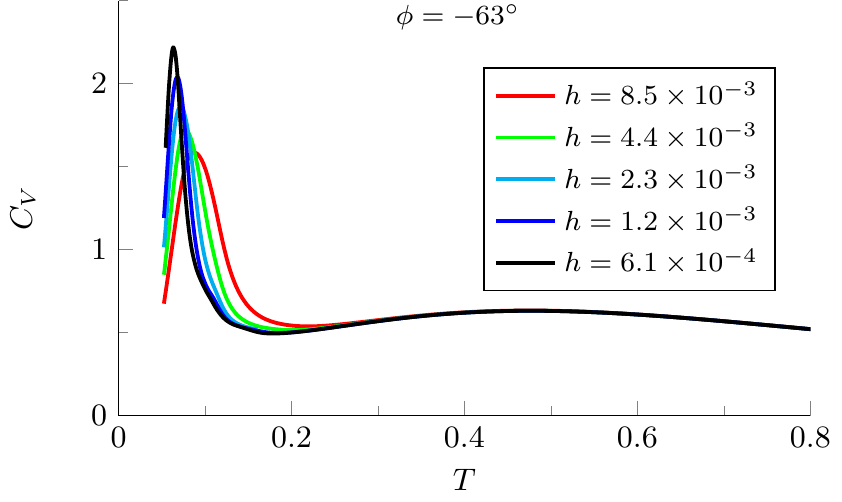}
\vspace{-0cm}
\caption{
The specific heat $C_V$ at $\phi = -63^{\circ}$ for small biases $h$ and $D=8$. At large $T$, comparable to the coupling constants in the Hamiltonian, there is a broad maximum that does not depend on $h$. At low $T$ there are sharp peaks moving towards lower $T$ with decreasing $h$. They indicate the symmetry breaking phase transition.
} 
\label{fig:cvpeaks}
\end{figure}

\begin{figure}[ht!]
\includegraphics[width=\columnwidth,clip=true]{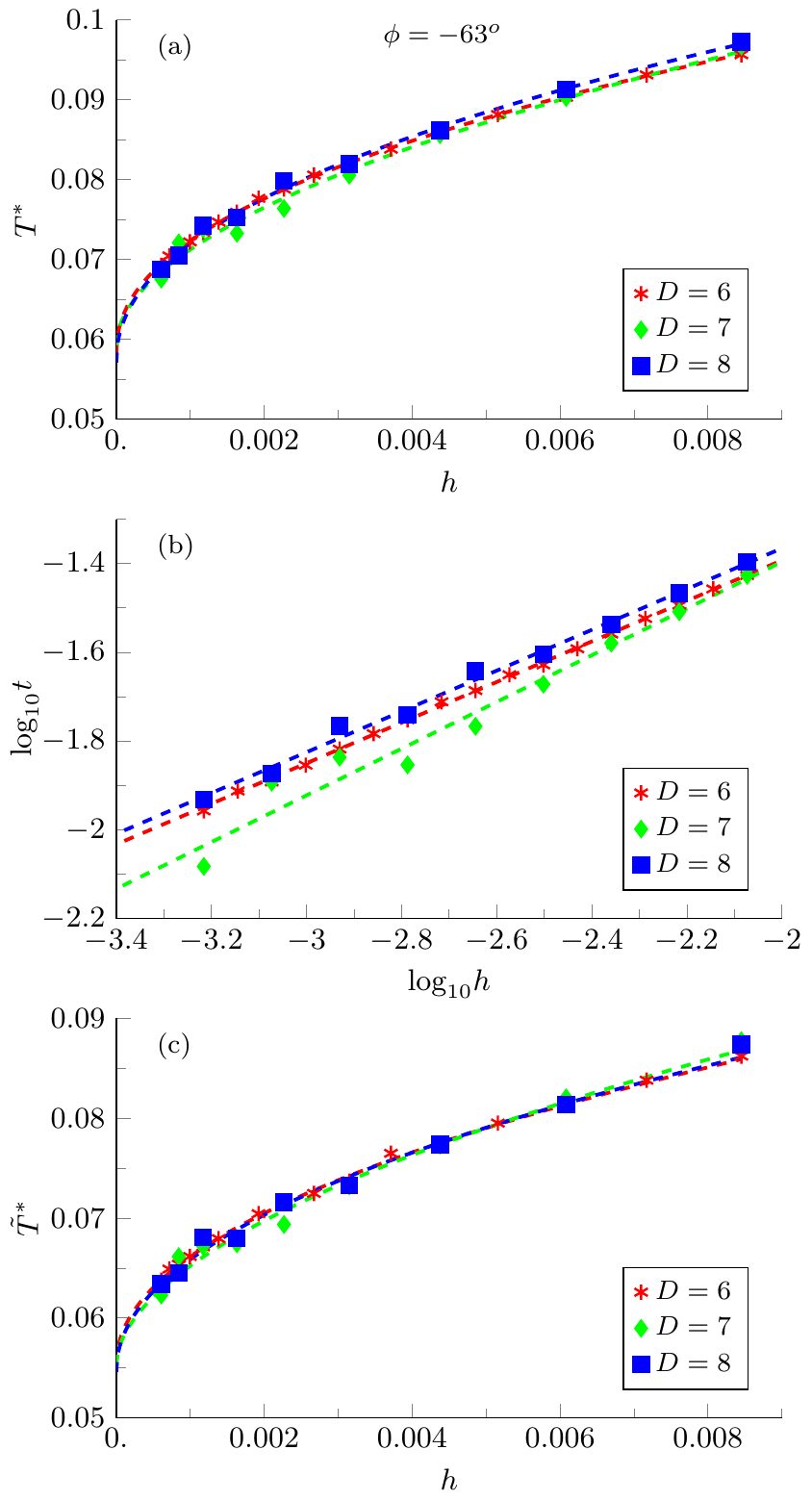}
\vspace{-0cm}
\caption{
In (a),
the pseudo-critical temperature $T^*$ at $\phi=-63^\circ$, obtained from the peaks of  $m'$, in function of the bias $h$ for bond dimensions $D=6,7,8$. The numerical data (points) are fitted with good accuracy by the scaling ansatz (\ref{tpeak}) (solid lines). 
In (b), 
a log-log plot of $t=T^* - T_c$ vs. $h$ demonstrates the scaling behavior. Here $T_c$ and the lines are the best fits obtained in panel (a). 
In (c), 
the same scaling analysis as in panel (a), but for $T^*$ obtained from the low-$T$ peaks of the specific heat $C_V$ in Fig. \ref{fig:cvpeaks}.
The mutually consistent values of $T_c$ and $1/\beta\delta$ fitted in (a) and (c) are collected in Tab. \ref{tab:th-63}.
} 
\label{fig:th-63}
\end{figure}

\begin{table}[h]
\begin{tabular}{|c|c|c|c|l|l|}
\hline
the method & $D$ & $\chi$ & d$\beta$ & $T_c$ & $1/\tilde\beta\delta$ \\ 
\hline
$m'$ peaks  & $6$ & $42$ & $0.01$  & $0.058(2)$ & $0.46(4)$ \\
$m'$ peaks & $7$ & $21$ & $0.01$  & $0.059(14)$ & $0.53(36)$  \\
$m'$ peaks & $8$ & $24$ & $0.01$  & $0.057(9)$ & $0.46(18)$  \\
$m'$ peaks & $6$ & $24$ & $0.005$  & $0.059(2)$ & $0.48(4)$ \\
$m'$ peaks & $6$ & $24$ & $0.02$  & $0.058(4)$ & $0.45(9)$ \\
$m'$ peaks & $6$ & $18$ & $0.01$  & $0.058(8)$ & $0.46(16)$ \\
$m'$ peaks & $6$ & $30$ & $0.01$  & $0.058(4)$ & $0.45(9)$ \\
$C_V$ peaks & $6$ & $42$ & $0.01$  & $0.055(3)$ & $0.48(8)$ \\
$C_V$ peaks & $7$ & $21$ & $0.01$  & $0.055(11)$ & $0.54(36)$  \\
$C_V$ peaks & $8$ & $24$ & $0.01$  & $0.055(11)$ & $0.48(30)$  \\
\hline
\end{tabular}
\caption{The critical temperature $T_c$ and the exponent $1/\tilde\beta\delta$ at $\phi=-63^{\circ}$ obtained by fitting $T^*(h)$ with the scaling ansatz (\ref{tpeak}), see Fig. \ref{fig:th-63}. Here we gather results obtained by the two methods of estimating $T^*(h)$, using to that end location of peaks either in $m'(t,h)$ or $C_V(t,h)$. The errors are statistical uncertainties of the fits for a confidence level $95\%$. The results for different choices of $\chi$ and $d\beta$ suggests that the modest $\chi=3D$ and $d\beta = 0.01$ are sufficient to obtain converged estimates of $T_c$ and $1/\tilde\beta\delta$. 
Furthermore, both methods of $T^*$-estimation give mutually consistent results. Combining all the results we finally estimate $T_c = 0.057(2)$, $1/\beta\delta = 0.46(5)$. To obtain the final estimates we use the weighted average, taking the error as the statistical error of the  average for a confidence level $99.7\%$.    }
\label{tab:th-63}
\end{table}

\section{Results}
\label{results}

We choose to study two angles: $\phi=-63^\circ$ and $-17^\circ$.  The former sits midway between the zero-temperature phase boundary at $-80^{\circ}$, separating the stripy phase from the spin liquid, and the $\textrm{SU}(2)$-symmetric point at $-45^{\circ}$. Likewise, the latter sits midway between the stripy-AF phase transition and the $SU(2)$-symmetric Heisenberg point at $0^\circ$. This is why we expect a relatively high critical temperature at both angles. Furthermore, $\phi=-63^\circ$ lies near the range $J/K = -0.3...-0.1$ reported recently \cite{ValentiReview} for a proximate Kitaev spin liquid material $\alpha\textrm{-RuCl}_3$, making it a good starting point for a future study of more realistic extensions of the minimal KH model.

The duality transformation (\ref{duality}) maps the results for $\phi=-63^\circ$ and $-17^\circ$ to $\phi=-136.09^\circ$ in the ferromagnetic phase and $145.23^\circ$ in the zig-zag phase, respectively. 

\subsection{Stripy (ferromagnetic) phase:\\
$\phi=-63^\circ$ ($\phi=-136.09^\circ)$}

The duality transformation (\ref{duality}) maps  $\phi=-63^\circ$ in the stripy phase to $\tilde\phi=-136.09^\circ$ in the ferromagnetic phase where we actually make simulations taking advantage of the fact that we need only two sublattices there. After the transformation the nearest neighbor terms in the Hamiltonian become
\be 
\tilde H_{i,j}^{\gamma} = 
\tilde A\left[ 
\cos(\tilde\phi)  \mathbf{  S}_i\cdot \mathbf{  S}_j  + 
2\sin(\tilde\phi)  S_i^{(\gamma)}  S_j^{(\gamma)}
\right].
\ee 
where $\tilde A = - A\cos(\phi)/\cos(\tilde \phi)$. $\tilde A \approx 0.63$ when we set $A=1$. 
The order parameter for the stripy phase equals the ferromagnetic order parameter of the transformed Hamiltonian 
$
m = 2\sqrt{\langle  S^x\rangle^2 +\langle  S^y\rangle^2 +  \langle  S^z\rangle^2}.
$ 

The specific heat for different values of the small bias $h$ is shown in Fig.~\ref{fig:cvpeaks}. At higher temperatures, comparable to the coupling constants in the Hamiltonian, we obtain a broad peak which does not depend on the bias. At an order of magnitude lower temperatures we can see sharp peaks whose position and magnitude depend strongly on the applied bias. This sensitivity suggests that they indicate spontaneous symmetry breaking in the direction of the bias. Below we analyze this low temperature regime in detail employing numerical data obtained with bond dimensions $D=6,7,8$ for biases in the range $6.1 \times 10^{-4} \le h \le 8.5 \times 10^{-3}$. This is where the results appear converged in $D$. 

Fig. \ref{fig:th-63}(a,b) shows the pseudo-critical temperatures $T^*(h)$ obtained from the peaks of $m'$. The data are fitted accurately by the scaling ansatz (\ref{tpeak}), indicating a second-order phase transition. Fig.~\ref{fig:th-63}(c) shows the same for the pseudo-critical temperatures $\tilde{T}^*(h)$ obtained from the peaks of $C_V$. The results are again fitted accurately by the scaling ansatz (\ref{tpeak}). In  Tab. \ref{tab:th-63} we collect $T_c$ and $1/\tilde\beta\delta$ fitted in Fig.~\ref{tab:th-63}. We find that the results for different $D$ are mutually consistent. We remark that the results obtained with $D=7,8$ are more "noisy" than $D=6$, which may be related to stability issues in the full update\cite{fu,HubigCirac,Hasik}. Furthermore, the results obtained from the $m'$ and $C_V$ peaks are also mutually consistent. 
Finally, for $\phi=-63^{\circ}$ we  obtain  
\be 
T_c=0.057(2), \quad  1/\tilde\beta\delta = 0.46(5),
\ee 
see details in Tab.~\ref{tab:th-63}. Using the duality transformation, we obtain for $\phi=-136.09^{\circ}$
\be
T_c=0.091(4), \quad  1/\tilde\beta\delta = 0.46(5).
\ee

\begin{figure}[h]
\includegraphics[width=\columnwidth,clip=true]{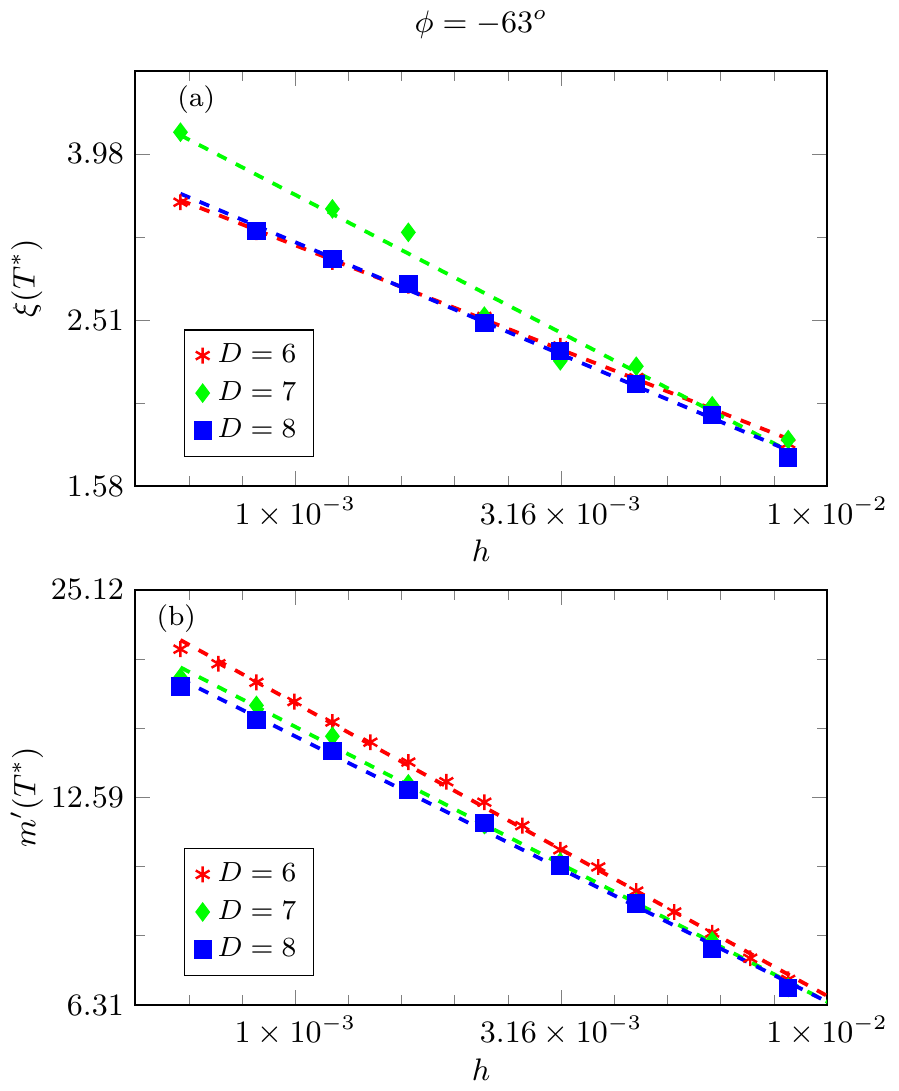}
\vspace{-0cm}
\caption{
Behavior of the correlation length $\xi$ and the order parameter's derivative $m'$ at the pseudo-critical temperature $T^{*}$ for $\phi=-63^{\circ}$.  In (a), a log-log plot of $\xi(T^*)$ (in units of the lattice constant of the rhombic lattice in Fig. \ref{fig:ctm}(b)) in function of the bias $h$ is fitted by the scaling ansatz (\ref{xipeak}) (dashed lines). In (b), a similar log-log plot of $m'(T^*)$ is fitted by the ansatz (\ref{m'peak}). Here $D$, $\chi$, $d\beta$, and the range of $h$ are the same as in Fig.~\ref{fig:th-63}. The critical exponents obtained from the fits are collected in Tab. \ref{tab:peaks}. In both cases the results are fitted by the scaling ansatzes with good accuracy. 
}     
\label{fig:peaks}
\end{figure}

\begin{table}
\begin{tabular}{|c|c|}
\hline
$D$ & the exponent\\ 
\hline
$6$ &   $\nu/\tilde\beta\delta=0.25(1)$ \\
$7$ &  $\nu/\tilde\beta\delta=0.33(5)$ \\
$8$ &  $\nu/\tilde\beta\delta=0.27(2)$ \\
$6$ &  $(\tilde\beta-1)/\tilde\beta\delta=0.423(8)$ \\
$7$ &  $(\tilde\beta-1)/\tilde\beta\delta=0.40(2)$ \\
$8$ &   $(\tilde\beta-1)/\tilde\beta\delta=0.38(2)$ \\
\hline
\end{tabular}
\caption{
The exponents $\nu/\tilde\beta\delta$ and $(\tilde\beta-1)/\tilde\beta\delta$ obtained by fitting $\xi(T^*,h)$ and $m'(T^*,h)$ in Fig. \ref{fig:peaks}.
}
\label{tab:peaks}
\end{table}

As a further self-consistency check, we analyze the correlation length $\xi(t^*,h)$ at $T^*$. We extract $\xi$ from the iPEPS with the precise method in Ref. \onlinecite{Rams_xiD_18}. Figure \ref{fig:peaks}(a) shows a log-log plot of $\xi(t^*,h)$ in function of $h$, which is fitted well by a linear behaviour predicted by the scaling ansatz (\ref{xipeak}). The fits for different $D$ yield close to each other values of the exponent $\nu/\tilde\beta\delta$, see Tab. \ref{tab:peaks}. A weighted average combining the results for different $D$ yields
\be
\nu/\tilde\beta\delta = 0.41(1).
\ee
Notice that the largest $\xi(t^*,h)$ of $3.5$ lattice constants of the rhombic lattice obtained in Fig. \ref{fig:ctm}(b) is beyond reach of the state of the art finite cluster exact diagonalization or DMRG on a cylinder. We analyze also $m'(t^*,h)$, see Fig. \ref{fig:peaks}(b). Again we find that the results can be accurately fitted by the scaling ansatz (\ref{m'peak}). The estimates of $(\tilde\beta-1)/\tilde\beta\delta$ obtained with different $D$ are close to each other, see  Tab. \ref{tab:peaks}. Their combination yields 
\be
(\tilde\beta-1)/\tilde\beta\delta = 0.26(1).
\ee

\begin{figure}[h]
\includegraphics[width=\columnwidth,clip=true]{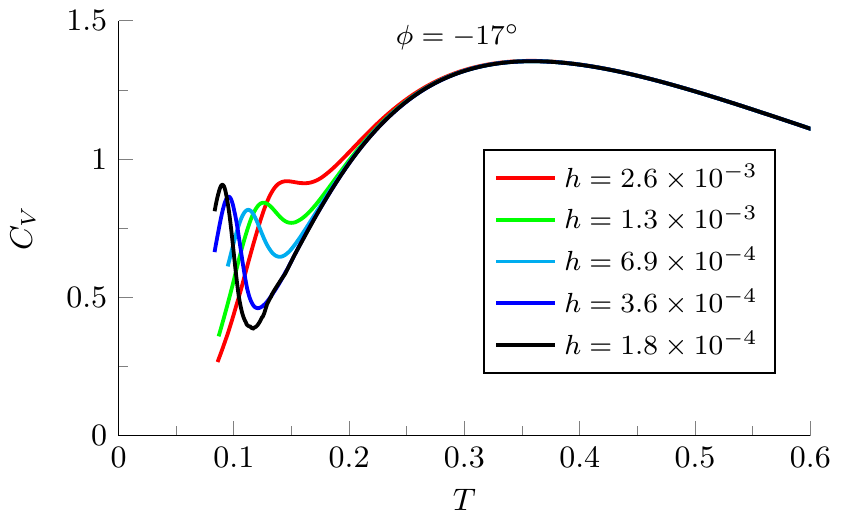}
\vspace{-0cm}
\caption{
$C_V$ peaks for $\phi = -17^{\circ}$ obtained with $D=8$ for different biases $h$. 
There is a broad maximum at high temperatures, that does not depend on the bias,
and sharp peaks at low temperatures whose location depends on $h$.
} 
\label{fig:cv-17}
\end{figure}

\begin{figure}[t]
\includegraphics[width=\columnwidth,clip=true]{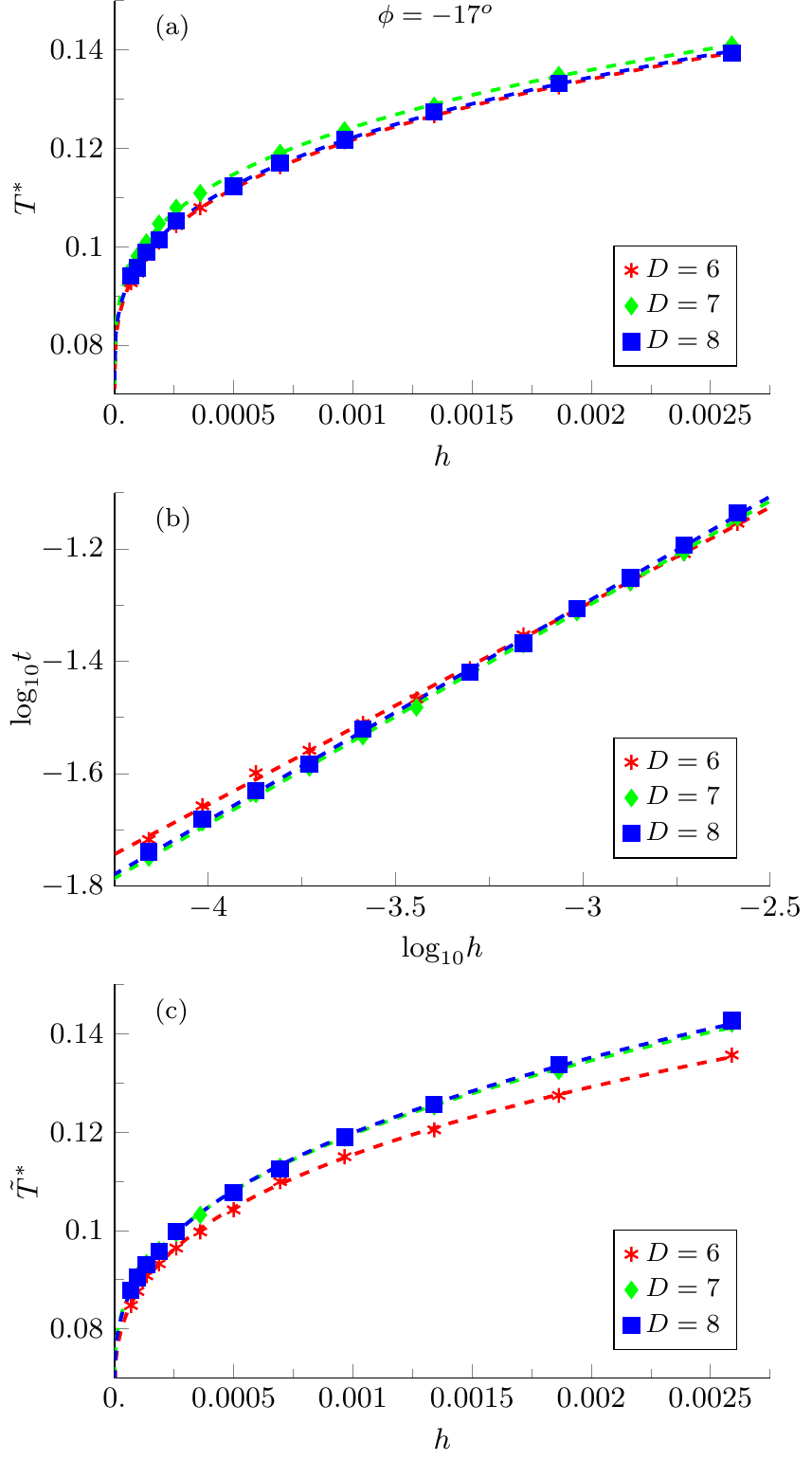}
\vspace{-0cm}
\caption{ 
In (a),  
$T^*$ obtained from the peaks of $m'$ at $\phi=-17^\circ$ in function of the bias $h$.
The numerical data (points) are fitted with the scaling ansatz  (\ref{tpeak}) (dashed lines) for $D=6,7,8$. Parameters of the fits are listed in Tab. \ref{tab:th-17}. 
In (b),  
the same results as in (a) plotted as a log-log plot of $t = T^{*}-T_c$ in function of $h$. 
Here the dashed lines are the best fits obtained in panel (a). 
In (c), 
the same scaling analysis as in panel (a), but for $\tilde{T}^*$ obtained from the low-$T$ peaks of the specific heat $C_V$ in Fig. \ref{fig:cv-17}.}     
\label{fig:th-17}
\end{figure}

\begin{table}[t]
\begin{tabular}{|c|c|c|c|l|l|}
\hline
method &  $D$ & $T_c$ & $1/\tilde\beta\delta$ \\ 
\hline
$m'$ peaks &  $6$ &  $0.071(4)$ & $0.31(3)$ \\
$C_V$ peaks &  $6$ &  $0.066(4)$ & $0.35(4)$ \\
$m'$ peaks &  $7$ &  $0.072(6)$ & $0.29(4)$ \\
$C_V$ peaks &  $7$ &  $0.070(4)$ & $0.38(4)$ \\
$m'$ peaks &  $8$ &  $0.073(4)$ & $0.32(2)$ \\
$C_V$ peaks &  $8$ &  $0.070(4)$ & $0.38(4)$ \\
\hline
\end{tabular}
\caption{ 
The critical temperatures $T_c$ and the critical exponents $1/\tilde\beta\delta$ obtained for $\phi=-17^\circ$ in Figs. \ref{fig:th-17}. The $T_c$ estimates agree within the uncertainties. }
\label{tab:th-17}
\end{table}

\begin{figure}[htb]
\includegraphics[width=\columnwidth,clip=true]{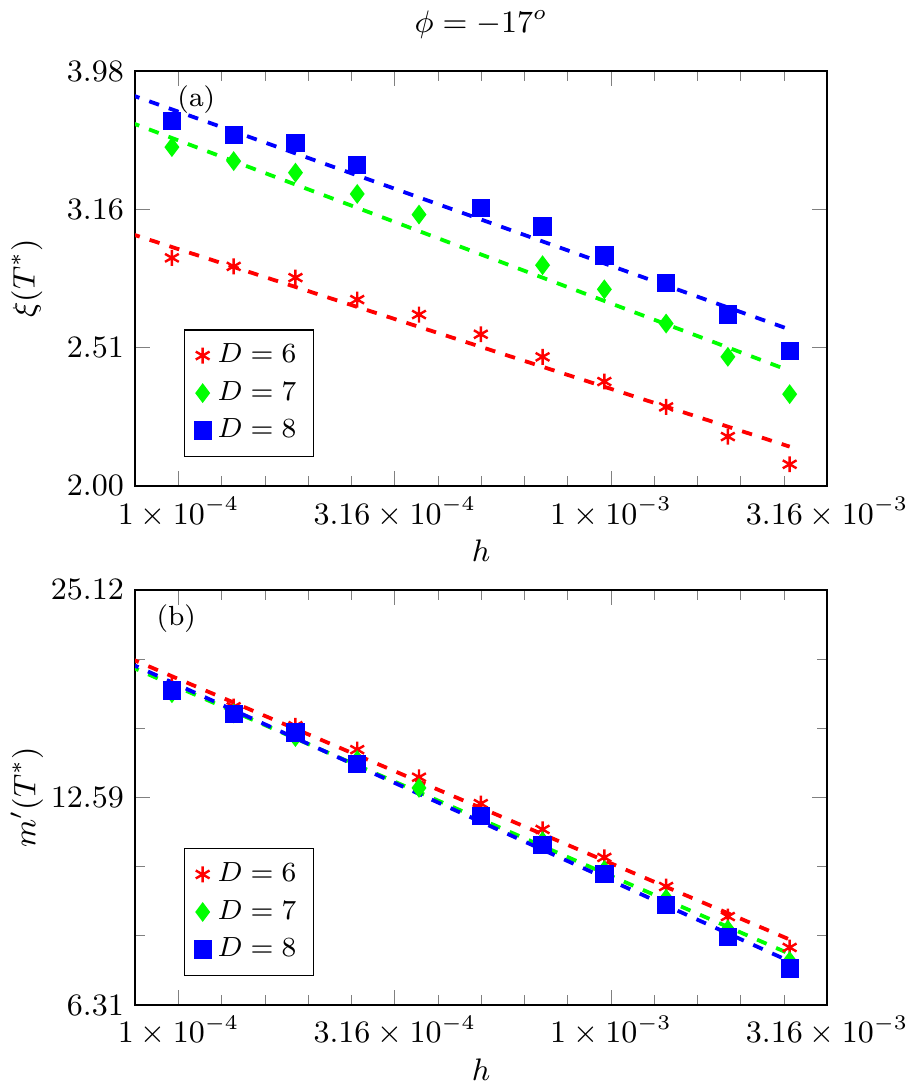}
\vspace{-0cm}
\caption{
In (a),
$\xi(t^*,h)$ for $D=6,7,8$ fitted by the scaling ansatz (\ref{xipeak}). The dashed lines are the best fits.
In (b),
$m'(t^*,h)$ fitted by the scaling ansatz (\ref{m'peak}). 
}     
\label{fig:xih-17}
\end{figure}

\subsection{Antiferromagnetic (zig-zag)  phase: \\
$\phi=-17^\circ$ ($\phi=145.23^{\circ}$)}

The order parameter is the staggered magnetization 
$m=\sqrt{\langle S^x_A-S^x_B\rangle^2 +\langle  S^y_A-S^y_B\rangle^2 +  \langle  S^z_A-S^z_B\rangle^2}$. In the following we present results for the bond dimensions $D=6,7,8$ and the time step $d\beta=0.01$.
We begin with Fig. \ref{fig:cv-17} showing the specific heat in function of temperature. Similarly as at $\phi=-63^\circ$, there is a broad maximum at temperatures comparable to the coupling constants in the Hamiltonian that, in the limit of small biases, does not depend on the applied bias. Therefore, it  is not related to symmetry breaking in contrast to the sharp peaks at low temperatures that depend on the bias in a systematic way. We investigate the peaks in detail below finding results consistent with a continuous, symmetry breaking, phase transition. 

In Fig. \ref{fig:th-17} we use $T^*$ obtained from the peaks of $m'$ and $C_V$ to estimate the critical temperature and the exponents. Their best fits are collected in table \ref{tab:th-17}. Averaging the results for $D=8$ we obtain:
\be 
T_c=0.072(3), \quad 1/\tilde\beta\delta=0.35(2).
\ee  
Using the duality (\ref{duality}) we obtain for $\phi=145.23^{\circ}$ in the zigzag phase:
\be 
T_c=0.062(3), \quad 1/\tilde\beta\delta=0.35(2).
\ee  

For a further self-consistency check, in the log-log plots in Fig. \ref{fig:xih-17} we test the scaling ansatzes (\ref{m'peak},\ref{xipeak}). We see that for $\xi(t^*,h)$ deviations from the power law are significant. Furthermore, the range of $\xi(t^*,h)$ is more limited here than for $\phi =-63^{\circ}$. For the maximal slope $m'(t^*,h)$ the deviations from the power law scaling are less significant than for $\xi(t^*,h)$.

The results for $\phi=-17^\circ$ are more significantly affected by the deviations from the asymptotic scaling than the ones for $\phi=-63^\circ$. Nevertheless, they provide evidence that $T_c$ is small w. r. t. the couplings in the Hamiltonian.

\begin{figure}[h]
\includegraphics[width=0.99\columnwidth,clip=true]{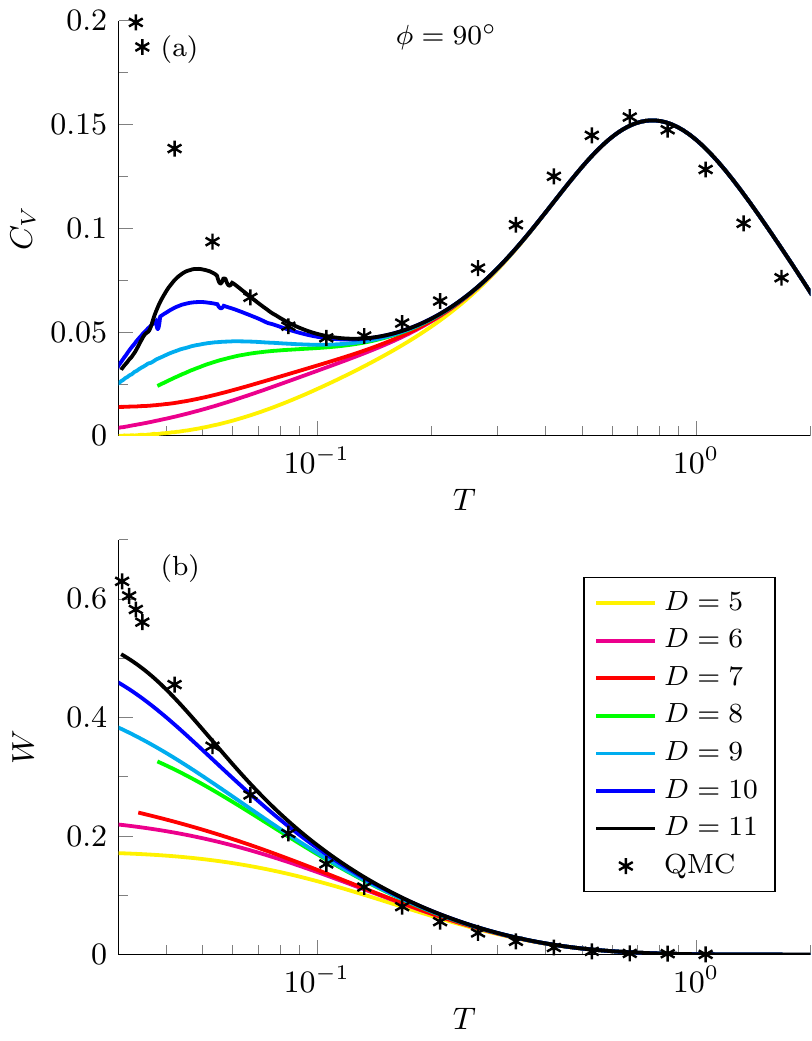}
\vspace{-0cm}
\caption{
In (a),
the specific heat $C_V$ in function of temperature in the Kitaev model for different bond dimensions $D$.
We also show quantum Monte Carlo data (QMC) from Ref. \onlinecite{Nasu2D}.
The broad peak at high temperature is well converged in $D$. 
Its small deviation from the QMC data may be due to a finite cluster size in the QMC calculations.
In (b),
the expectation value of the flux operator in function of temperature.
}     
\label{fig:KHCV}
\end{figure}

\subsection{Kitaev quantum spin liquid: $\phi=90^\circ$}
\label{secKitaev}

The pure Kitaev model is tractable by quantum Monte-Carlo \cite{Nasu2D}. At the same time its critical spin-liquid ground state makes the low temperature physics challenging for tensor networks as they may require large bond dimension $D$. Therefore, it is an ideal case to test limitations of our method by benchmarking it against the quantum Monte-Carlo results. The model was shown \cite{Nasu2D} to have two cross-over temperatures: $T_1<T_2$. Near $T_2$ it crosses over from a spin disordered paramagnet to a state with NN spin correlations. Near $T_1$ flux ordering takes place. This is where an expectation value of the plaquette flux operator, $W$, becomes non-zero before it converges to $1$ in the ground state at $T=0$.

In the absence of any phase transitions at finite temperature, there is no need to smooth the evolution
by any bias field. Figure \ref{fig:KHCV}(a) shows specific heat in function of temperature. We  can clearly see the peak at $T_2$ which is well converged in the bond dimension $D$. The second peak at the lower $T_1$ builds up with increasing $D$ in a systematic way. Location of the peaks is consistent with Ref. \onlinecite{Nasu2D}. The origin of the second peak is corroborated in Fig. \ref{fig:KHCV}(b) showing the average flux operator $W$ in function of temperature. 

\section{Conclusion}
\label{conclusion}

We applied the recently introduced tensor network algorithm to obtain thermal states of the Kitaev-Heisenberg model with a focus on their critical properties. 
As a technical advancement,
we also show that the dynamical mapping from the hexagonal to the rhombic lattice makes the exact environment full update (eeFU) as efficient as the simple full update (FU) algorithm where the infinite tensor environment is delayed with respect to the imaginary time evolution.
In the stripy phase at $\phi=-63^\circ$ we provide evidence for the second order phase transition and estimate its critical temperature at $T_c=0.056(4)$. 
Furthermore, for $\phi=-17^{\circ}$ in the antiferromagnetic phase, we estimate $T_c=0.076(15)$. Both critical temperatures are small w. r. t. the couplings in the Hamiltonian.
By the duality transformation, these results can be mapped to, respectively, ferromagnetic and zigzag phases. 
Finally,
we benchmark our method against quantum Monte-Carlo results in the special case of pure Kitaev model with
a challenging spin-liquid ground state. We recover two crossovers for spin and flux ordering.


\acknowledgments

We acknowledge insightful discussions with Andrzej M. Ole\'s on the KH model. 
We thank Joji Nasu for the QMC data in the Kitaev limit.  
This research was supported in part by the Polish Ministry of Science and Education under 
grant DI2015 021345 (AF) and the National Science Centre (Narodowe Centrum Nauki) under grant 2016/23/B/ST3/00830 (PC,AF,JD).


\bibliographystyle{apsrev4-1}
\bibliography{KH.bib}

\end{document}